\documentclass[10pt]{article}
\usepackage{amsmath}
\usepackage{amsfonts}
\usepackage{amssymb}

\begin{document}

\author{J\'{e}r\^{o}me Perez$^{(1),(2)}$\\$^{(1)}$Ecole Nationale Sup\'{e}rieure de Techniques Avanc\'{e}es\\Laboratoire de Math\'{e}matiques Appliqu\'{e}es\\32\ Boulevard Victor, 75015\ Paris, France\\$^{(2)}$Laboratoire de l'Univers et de se TH\'{e}ories, CNRS UMR\ 8631,\\Observatoire de Paris-Meudon\\5 place Jules Janssen, 92195 Meudon, France}
\title{Vlavov and Poisson equations in the context of self-gravitating systems}
\maketitle
\begin{abstract}
We present Vlasov's equation and its association with Poisson's equation in the context of modelling self-gravitating systems such as Globular Clusters or Galaxies.
We first review the classical hypotheses of the model. We continue with a presentation of the Hamilton-Morrison structure of Vlasov's equation to study the equilibrium and the stability  of self-gravitating systems. Finally, we present some preliminary results concerning some properties of the time dependent solutions of the Vlasov-Poisson system.
\end{abstract}

\section{Hypotheses}

\bigskip It is well known that Vlasov's equation can be used to describe the
dynamics of a self gravitating system. We recall the usual descriptive hypotheses.

First, we consider a set of $N$ point mass particles of
individual masses $m_{i=1,..,N}$ and we consider the associated $N$ particle
distribution function in the phase space $f^{\left(  N\right)  }$ such that, at
any time $t$
\begin{equation}
1=%
{\displaystyle\int}
f^{\left(  N\right)  }\left(  \mathbf{\Gamma}_{1},\cdots,\mathbf{\Gamma}%
_{N},t\right)  \,\prod_{i=1}^{N}d\mathbf{\Gamma}_{i}\;\;\; ,\label{un}%
\end{equation}
where $\mathbf{\Gamma}_{i}=\left(  \mathbf{p}_{i},\mathbf{q}_{i}\right)  $ is
the 6-vector which indicates the position $\mathbf{q}_{i}$ and the impulsion
$\mathbf{p}_{i}$ of the $i^{th}$ particle. Provided that each $\mathbf{p}_{i}$
and $\mathbf{q}_{i}$ are Hamiltonian conjugated variables and all the forces
are conservative, one can obtain directly from this normalization equation a
conservation equation (Liouville's theorem)
\begin{equation}
\frac{\partial f^{\left(  N\right)  }}{\partial t}+\sum_{i=1}^{N}\left(
\frac{\mathbf{p}_{i}}{m}.\frac{\partial f^{\left(  N\right)  }}{\partial
\mathbf{q}_{i}}-\frac{\partial U_{i}}{\partial\mathbf{q}_{i}}.\frac{\partial
f^{\left(  N\right)  }}{\partial\mathbf{p}}\right)  =0
\end{equation}
where $U_{i}$ is the potential energy at the origin of the force
$\mathbf{f}_{i}$ seen by the $i^{th}$ particle
\begin{equation}
\mathbf{f}_{i}=\frac{d\mathbf{p}_{i}}{dt}=-\frac{\partial U_{i}}%
{\partial\mathbf{q}_{i}}%
\end{equation}
If the only force considered is gravitation, we have
\begin{equation}
U_{i}=-G\sum_{i\neq j=1}^{N}\frac{m_{i}m_{j}}{\left|  \mathbf{q}%
_{i}-\mathbf{q}_{j}\right|  }%
\end{equation}
In order to obtain a gravitational Vlasov's equation, we first need to introduce
marginal distribution functions $f^{\left(  1\right)  }$,and $f^{\left(  2\right)
}$ respectively for one and two particles as
\begin{align}
f^{\left(  1\right)  }\left(  \mathbf{\Gamma}_{i},t\right)    & =\int
f^{\left(  N\right)  }\left(  \mathbf{\Gamma}_{1},\cdots,\mathbf{\Gamma}%
_{N},t\right)  \,\prod_{j\neq i=1}^{N}d\mathbf{\Gamma}_{j}\\
f^{\left(  2\right)  }\left(  \mathbf{\Gamma}_{i},\mathbf{\Gamma}%
_{j},t\right)    & =\int f^{\left(  N\right)  }\left(  \mathbf{\Gamma}%
_{1},\cdots,\mathbf{\Gamma}_{N},t\right)  \,\prod_{k\neq j\neq i=1}%
^{N}d\mathbf{\Gamma}_{k}%
\end{align}
We then make use of the Boltzmann ansatz which assumes the existence of a
two-particle correlation function $k\left(  \Gamma_{i},\Gamma_{j},t\right)  $
such that
\begin{equation}
\forall i\neq j=1,..,N\qquad f^{\left(  2\right)  }\left(  \mathbf{\Gamma}%
_{i},\mathbf{\Gamma}_{j},t\right)  =f^{\left(  1\right)  }\left(
\mathbf{\Gamma}_{i},t\right)  f^{\left(  1\right)  }\left(  \mathbf{\Gamma
}_{j},t\right)  +k\left(  \mathbf{\Gamma}_{i},\mathbf{\Gamma}_{j},t\right)
\end{equation}
We now have suppose that the particles are indiscernible\footnote{In
particular\ this means that all particles have the same mass $m$.}, and have
the same probability law $f\left(  \mathbf{\Gamma},t\right)  =f^{\left(
1\right)  }\left(  \mathbf{\Gamma},t\right)  /N$. Some direct calculations
then give
\begin{equation}
\frac{\partial f}{\partial t}+\frac{\mathbf{p}}{m}.\frac{\partial f}%
{\partial\mathbf{q}}-m\frac{\partial\psi}{\partial\mathbf{q}}.\frac{\partial
f}{\partial\mathbf{p}}=N^{2}Gm%
{\displaystyle\int}
\frac{\partial k\left(  \mathbf{\Gamma},\mathbf{\Gamma}^{\prime},t\right)
}{\partial\mathbf{p}}\frac{\left(  \mathbf{q}-\mathbf{q}^{\prime}\right)
}{\left|  \mathbf{q}-\mathbf{q}^{\prime}\right|  ^{3}}d\mathbf{\Gamma}%
^{\prime}%
\end{equation}
where $\psi\left(  \mathbf{q},t\right)  $ is the mean field gravitational
potential defined from $f$ by the inverse Poisson equation
\begin{equation}
\psi\left(  \mathbf{q},t\right)  =-Gm%
{\displaystyle\int}
\frac{f\left(  \mathbf{\Gamma}^{\prime},t\right)  }{\left|  \mathbf{q}%
-\mathbf{q}^{\prime}\right|  }d\mathbf{\Gamma}^{\prime}%
\end{equation}

It is well known $\left(  \text{see e.g. Binney and Tremaine\cite{bt}}\right)
$ that in the case of self-gravitating systems the time $T_{r}$ taken by
collisions to relax it, can be evaluated by the Chandrasekhar formula
\begin{equation}
T_{r}\approx\frac{N}{\ln N}T_{d}\approx\frac{N}{\ln N}\left(  G\overline{\rho
}\right)  ^{-1/2}%
\end{equation}
where $T_{d}$ is the dynamical or crossing time of the system and
$\overline{\rho}$ its mean mass density. The collisional relaxing time then
ranges from $10^{9,10}yr$ for Globular Clusters $\left(  N\sim10^{5}\right)  $
or $10^{11,15}yr$ for a typical galaxy $\left(  N\sim10^{10}\right)  $ to
$10^{13,17}yr$ for Galaxy Clusters $\left(  N\sim10^{3}\right)  .$ Hence,
during a reasonable time of about a few hundreds dynamical times one can
actually consider that two-particle correlations (namely gravitational
collisions) are dynamically unimportant, and by consequence that
\begin{equation}
k\left(  \mathbf{\Gamma},\mathbf{\Gamma}^{\prime},t\right)  \equiv0
\end{equation}
During this long period the mean field dynamics of a self-gravitating system
are governed by the Vlasov-Poisson system
\begin{align}
\frac{\partial f}{\partial t}+\frac{\mathbf{p}}{m}.\frac{\partial f}%
{\partial\mathbf{q}}-m\frac{\partial\psi}{\partial\mathbf{q}}.\frac{\partial
f}{\partial\mathbf{p}}  & =0\\
\psi\left(  \mathbf{q},t\right)    & =-Gm%
{\displaystyle\int}
\frac{f\left(  \mathbf{\Gamma}^{\prime},t\right)  }{\left|  \mathbf{q}%
-\mathbf{q}^{\prime}\right|  }d\mathbf{\Gamma}^{\prime}%
\end{align}

\section{Vlasov-Poisson as a Hamiltonian system}

It is often quoted that Vlasov's equation is a Hamiltonian system. It is true
but the formulation of such a problem is generally unclear. Using classical
Poisson brackets defined for any scalar fields $A\left(  \mathbf{q}%
,\mathbf{p},t\right)  $ and $B\left(  \mathbf{q},\mathbf{p},t\right)  $ by
\begin{equation}
\left\{  A,B\right\}  :=\frac{\partial A}{\partial\mathbf{q}}.\frac{\partial
B}{\partial\mathbf{p}}-\frac{\partial B}{\partial\mathbf{q}}.\frac{\partial
A}{\partial\mathbf{p}}%
\end{equation}
one can write Vlasov's equation in the form%

\begin{equation}
\frac{\partial f}{\partial t}+\frac{\mathbf{p}}{m}.\frac{\partial f}%
{\partial\mathbf{q}}-m\frac{\partial\psi}{\partial\mathbf{q}}.\frac{\partial
f}{\partial\mathbf{p}}=\frac{\partial f}{\partial t}+\left\{  f,E\right\}
=0\label{vlacro}%
\end{equation}
where
\begin{equation}
E=\frac{\mathbf{p}^{2}}{2m}+m\psi=\frac{\mathbf{p}^{2}}{2m}-Gm^{2}%
{\displaystyle\int}
\frac{f\left(  \mathbf{\Gamma}^{\prime},t\right)  }{\left|  \mathbf{q}%
-\mathbf{q}^{\prime}\right|  }d\mathbf{\Gamma}^{\prime}\label{enepp}%
\end{equation}
is the mean field energy per particle. It is a fact that relation $\left(
\ref{vlacro}\right)  $ is not a Hamiltonian equation. As a matter of fact,
general Hamiltonian equations give the \emph{total} derivative of a canonical
variable from a lie bracket between this variable and a
propagator.\ Nevertheless, a series of works initiated by
Morrison\cite{morrison80} , have shown that a Hamiltonian formulation of non
dissipative physics is generally possible.\ Let us recall these results in the
context of Vlasov's equation.

We consider any physical functional of the distribution function, in a very
general context this functional can be written
\begin{equation}
F\left[  f\right]  =\int\varphi\left(  f,\mathbf{\Gamma}\right)
d\mathbf{\Gamma}\label{genf}%
\end{equation}
where $\varphi$ is a very general suitable function, for example if $\varphi$
do not depend on $\mathbf{\Gamma}$, $F\left[  f\right]  $ is called a Casimir
and it can be easily proven that this is a conserved quantity during the
Vlasov dynamics. By direct calculation we have
\begin{equation}
\frac{dF}{dt}=\frac{\partial F}{\partial t}=\int\frac{\partial f}{\partial
t}\frac{\partial\varphi}{\partial f}d\mathbf{\Gamma}\label{dFdt}\;\;\;.
\end{equation}
It is important to note here that
\begin{equation}
\frac{\partial\varphi}{\partial f}=\frac{\delta F}{\delta f}\label{Dfdf}%
\end{equation}
where $\delta/\delta f$ indicates the usual functional derivative defined in
the case of $\left(  \ref{genf}\right)  $ by the relation
\begin{equation}
F\left[  f+\delta f\right]  -F\left[  f\right]  =%
{\displaystyle\int}
\left[  \frac{\delta F}{\delta f}\delta f+\mathcal{O}\left(  \delta f\right)
\right]  d\mathbf{\Gamma}\label{defder}%
\end{equation}
Using $\left(  \ref{dFdt}\right)  $, $\left(  \ref{Dfdf}\right)  $ and
$\left(  \ref{vlacro}\right)  ,$ one can then write
\begin{equation}
\frac{dF}{dt}=\int\left\{  E,f\right\}  \frac{\delta F}{\delta f}%
d\mathbf{\Gamma}\label{presque}%
\end{equation}
Using definition $\left(  \ref{defder}\right)  ,$ one can verify that
the total energy contained in the system
\begin{equation}
H\left[  f\right]  :=%
{\displaystyle\int}
\frac{\mathbf{p}^{2}}{2m}f\left(  \mathbf{\Gamma},t\right)  d\mathbf{\Gamma
}-\frac{Gm^{2}}{2}%
{\displaystyle\int}
\frac{f\left(  \mathbf{\Gamma},t\right)  \text{ }f\left(  \mathbf{\Gamma
}^{\prime},t\right)  }{\left|  \mathbf{q}-\mathbf{q}^{\prime}\right|
}d\mathbf{\Gamma}d\mathbf{\Gamma}^{\prime}%
\end{equation}
which is a functional of the form $\left(  \ref{genf}\right)  $, is such that
\begin{equation}
\frac{\delta H}{\delta f}=E\label{dHdf}\;\;\;.
\end{equation}
Then, integrating the \textsc{rhs} of $\left(  \ref{presque}\right)  $ by
parts\footnote{This is actually a integration by parts. One can directly show
indeed that for functions that decay sufficiently rapidly at $\left|
\mathbf{q}\right|  $ and $\left|  \mathbf{p}\right|  $ tend to $\infty$ we
have
\begin{equation}
\int d\mathbf{\Gamma}A\left\{  B,C\right\}  =-\int d\mathbf{\Gamma}C\left\{
B,A\right\}  =\int d\mathbf{\Gamma}\left\{  A,B\right\}  C
\end{equation}
In the case of equation $\left(  \ref{dret}\right)  $ surface terms vanish as
the distribution function vanishes at infinity.} and employing $\left(
\ref{dHdf}\right)  $ one gets
\begin{equation}
\frac{dF}{dt}=\int\left\{  \frac{\delta F}{\delta f},\frac{\delta H}{\delta
f}\right\}  fd\mathbf{\Gamma}\label{dret}%
\end{equation}
which is the result introduced, perhaps more directly, by
Morrison\cite{morrison80} . This last equation can be formulated as
\begin{equation}
\frac{dF}{dt}=\left\langle F,H\right\rangle \label{vlaham}%
\end{equation}
where we have introduced the Morrison's bracket defined for any functional
$A\left[  f\right]  $ and $B\left[  f\right]  $ of the form $\left(
\ref{genf}\right)  $, by the relation
\begin{equation}
\left\langle A,B\right\rangle =\int\left\{  \frac{\delta A}{\delta f}%
,\frac{\delta B}{\delta f}\right\}  fd\mathbf{\Gamma}\label{morbra}%
\end{equation}
Some very simple calculations show that Morrison's bracket is skew-symmetric,
verify Jacobi's identity and can be used to form a Lie algebra. Hence,
equation $\left(  \ref{vlaham}\right)  $ naturally defines the Hamiltonian
structure of Vlasov's equation. We can note that in this formulation there
is no conjugated variable to the canonical  $f$.

\section{\smallskip Some uses ...}

\subsection{Generalities}

It is well known that a regular Hamiltonian formulation of a problem provides room for much formal analysis. Let us see what is possible with Vlasov's
equation. In a pure formal way, one can always write
\begin{equation}
\frac{dF}{dt}=\left\langle F,H\right\rangle =-\left\langle H,.\right\rangle
F:=\mathcal{T}\left(  F\right)  \label{ODE} \;\;\;.
\end{equation}
Here, the functional operator $\mathcal{T}$ appearing in this last equation does not
depend explicitly on time\thinspace$t$, because the total energy of the system
$H\left[  f\right]  $ is a conserved quantity. Hence, one can utilize the
resolvant of the Ordinary Differential Equation $\left(  \ref{ODE}\right)  $
and we have
\begin{align}
F\left[  f\right]    & =\exp\left[  -\left(  t-t_{o}\right)  \left\langle
H,.\right\rangle \right]  F\left[  f_{o}\right]  \\
& =F\left[  f_{o}\right]  -\tfrac{\left(  t-t_{o}\right)  }{1!}\left\langle
H,F\left[  f_{o}\right]  \right\rangle _{f=f_{o}}+\tfrac{\left(
t-t_{o}\right)  ^{2}}{2!}\left\langle H,\left\langle H,F\left[f_{o}\right]\right\rangle \right\rangle _{f=f_{o}} \nonumber \\
& +\mathcal{O}\left(  \left(
t-t_{o}\right)  ^{2}\right)  
\end{align}
where $f_{o}=f\left(  \mathbf{\Gamma},t_{o}\right)  $. It is possible to
compare this expansion with the general Taylor expansion obtained from equation
$\left(  \ref{genf}\right)  $ which has the form
\begin{align}
F\left[  f\left(  \mathbf{\Gamma},t\right)  \right]    & =\int\varphi\left(
f\left(  \mathbf{\Gamma},t\right)  ,\mathbf{\Gamma}\right)  d\mathbf{\Gamma
}\\
& =%
{\displaystyle\int}
\left[  \varphi\left(  f\left(  \mathbf{\Gamma},t_{o}\right)  ,\mathbf{\Gamma
}\right)  +\tfrac{\left(  t-t_{o}\right)  }{1!}\frac{\partial\varphi\left(
f\left(  \mathbf{\Gamma},t_{o}\right)  ,\mathbf{\Gamma}\right)  }{\partial
f}\right.  \nonumber\\
& \qquad\left.  +\tfrac{\left(  t-t_{o}\right)  ^{2}}{2!}\frac{\partial
^{2}\varphi\left(  f\left(  \mathbf{\Gamma},t_{o}\right)  ,\mathbf{\Gamma
}\right)  }{\partial f^{2}}+\mathcal{O}\left(  \left(  t-t_{o}\right)
^{2}\right)  \right]  d\mathbf{\Gamma}%
\end{align}
Hence, we arrive at
\begin{align}
\frac{\partial\varphi\left(  f\left(  \mathbf{\Gamma},t_{o}\right)
,\mathbf{\Gamma}\right)  }{\partial f}  & =-\left\langle H,F\left[
f_{o}\right]  \right\rangle \\
\frac{\partial^{2}\varphi\left(  f\left(  \mathbf{\Gamma},t_{o}\right)
,\mathbf{\Gamma}\right)  }{\partial f^{2}}  & =\left\langle H,\left\langle
H,F\left[  f_{o}\right]  \right\rangle \right\rangle
\end{align}
etc ...
It is well known from quantum field theory that this expansion allows us see
that the total energy $H$ is the time propagator of the system. In a more
general way, Noether's Theorem says that any conserved quantity is associated
to a symmetry of the system, hence invoking a general canonical
transformation, the functional Vlasov's equation can be written (see Perez and
Lachieze-Rey\cite{perez96a} for details)
\begin{equation}
\frac{dF}{d\lambda}=\left\langle F,K\right\rangle \label{general}%
\end{equation}
where the pair $\left(  K\left[  f\right]  ,\lambda\right)  $ represents
 the conserved quantity (e.g. $H\left[  f\right]  $) and the
variable related to system's symmetry (e.g. $t-$translation), respectively.

Below, we present two natural uses of these seemingly formal developments

\subsection{Stability analysis}

We now consider a steady state solution $f_{o}\left(  \mathbf{\Gamma}\right)
$ of the Vlasov's equation. In a general way, Hamilton's theory of canonical
transformations says us that any physical perturbation of this equilibrium is
generated by some functional generator $G$, such that for any functional
$F\left[  f\right]  $ of the distribution function we have
\begin{equation}
\frac{dF}{d\varepsilon}=\left\langle F,G\right\rangle \label{hola}%
\end{equation}
where $\varepsilon$ controls the perturbation (see Perez and Lachieze-Rey
\cite{perez96a} and reference therein). In particular, if the perturbed state
is described by a distribution function $f\left(  \mathbf{\Gamma},t\right)  $
written as
\begin{equation}
f\left(  \mathbf{\Gamma},t\right)  =f_{o}\left(  \mathbf{\Gamma}\right)
+\varepsilon f_{1}\left(  \mathbf{\Gamma},t\right)  +\varepsilon^{2}%
f_{2}\left(  \mathbf{\Gamma},t\right)  +O\left(  \varepsilon^{2}\right)
\end{equation}
Taking $F=Id$ in equation $\left(  \ref{hola}\right)$, one can
then verify that
\begin{equation}
f_{1}\left(  \mathbf{\Gamma},t\right)  =\left\{  g,f_{o}\right\}  \text{,
\qquad}f_{2}\left(  \mathbf{\Gamma},t\right)  =\frac{1}{2!}\left\{  g,\left\{
g,f_{o}\right\}  \right\}  \text{, \qquad etc...}%
\end{equation}
where the so-called generating function $g$ is the functional derivative of
$G$. This kind of theory says that the linearized perturbation $f_{1}%
:=\left\{  g,f_{o}\right\}  $ is generated from the unperturbed distribution
$f_{o}\left(  \mathbf{\Gamma}\right)  $ via a canonical transformation by some
small perturbation $g\left(  \mathbf{\Gamma},t\right)  $.\ And so on for high
order non linear terms.

Considering such kinds of perturbation is not restrictive. As a matter of fact,
only perturbations of this form can be generated via a deformation that
enforces all the constraints associated with conservation of phase
(Liouville's theorem).

Taking $F=H$ in equation $\left(  \ref{hola}\right)  $ allows us to obtain
an energy perturbation induced by equilibrium perturbation%
\begin{align}
H\left[  f\right]    & =H_{o}+\varepsilon H_{1}+\varepsilon^{2}H_{2}%
+\mathcal{O}\left(  \varepsilon^{2}\right)  \label{expand}\\
& =H\left[  f_{o}\right]  -\varepsilon\left\langle G,H\right\rangle _{f=f_{o}%
}+\tfrac{1}{2}\varepsilon^{2}\left\langle G,\left\langle G,H\left[
f_{o}\right]  \right\rangle \right\rangle _{f=f_{o}}+\mathcal{O}\left(
\varepsilon^{2}\right)  \nonumber
\end{align}
The first two terms of this expansion are very useful in stability analysis.

\subsubsection{Equilibrium}

It is a well known property that if a system is in an equilibrium state, the
first order variation of its total energy vanishes. This property can be
directly shown from the use of the $\varepsilon-$order term in the expansion
$\left(  \ref{expand}\right)  $. As a matter of fact,
\begin{equation}
H_{1}=-\left\langle G,H\right\rangle _{f=f_{o}}=-\int\left\{  \frac{\delta
G}{\delta f},\frac{\delta H}{\delta f}\right\}  f_{o}d\mathbf{\Gamma=}\int
f_{o}\left\{  E,\frac{\delta G}{\delta f}\right\}  d\mathbf{\Gamma}%
\end{equation}
an integration by parts and the use of Vlasov's equation then gives
\begin{equation}
H_{1}=\int\left\{  f_{o},E\right\}  \frac{\delta G}{\delta f}d\mathbf{\Gamma
=}\int\frac{\partial f_{o}}{\partial t}\frac{\delta G}{\delta f}%
d\mathbf{\Gamma}%
\end{equation}
which is evidently a vanishing term if $f_{o}$ is an equilibrium state.

General equilibrium states are the following :
\begin{itemize}
\item $f_{o}=f_{o}\left(  E\right)  $ distribution functions represent all
$\mathbf{q}-$spherical and $\mathbf{p}-$isotropic equilibria of collisionless
self-gravitating systems ;

\item $f_{o}=f_{o}\left(  E,L^{2}\right)  $ distribution functions where
$L^{2}$ is the mean field squared angular momentum per particle, represent all
$\mathbf{q}-$spherical and $\mathbf{p}-$anisotropic equilibria of
collisionless self-gravitating systems ;

\item $f_{o}=f_{o}\left(  E,L_{z}\right)  $ distribution functions where
$L_{z}$ is the mean field $z-$component of the angular momentum per particle,
represent an equilibrium self-gravitating system with a $z-$symmetry in
$\mathbf{q}-$space.
\end{itemize}

For detailed proofs of these affirmations and properties of these equilibria,
we refer to Perez and Aly\cite{Perezaly} .

\subsubsection{Equilibrium stability}

The stability of a given equilibrium described by the time-independent
distribution function $f_{o}$ is classically related to the sign of the
$\varepsilon^{2}-$order term in the expansion $\left(  \ref{expand}\right)  $
\begin{equation}
H_{2}=\tfrac{1}{2}\left\langle G,\left\langle G,H\left[  f_{o}\right]
\right\rangle \right\rangle _{f=f_{o}}%
\end{equation}

More precisely, one can prove that an equilibrium state described by a
distribution function $f_{o}$ is guaranteed to be linearly stable if the
second variation $H_{2}$ is strictly positive for all perturbations. The
positivity of $H_{2}$ thus provides a sufficient criterion for linear
stability. If $H_{2}$ is of indeterminate sign, the situation is less
obvious. In fact the existence of negative energy perturbations does not necessarily imply an instability (there exist some
counterexample). However, one \emph{does} at least expect that it guarantees
non linear instability in the presence of dissipation. This idea goes back at
least to Moser~\cite{Moser} . The implications for the Vlasov's equation were
first considered by P.J.\ Morrison\cite{Morrison87} in the context of plasma
physics. That negative energy perturbations imply an instability towards
dissipation was proved more recently in a fairly general context by Bloch et
al~\cite{Blochetal} .

Let us indicates below the essential calculations which yields an explicit
formula for $H_{2}$. From $\left(  \ref{morbra}\right)  ,$ taking $g:=\delta
G/\delta f$, we get
\begin{equation}
H^{\left(  2\right)  }=\frac{1}{2}\int f_{o}\left\{  g,\frac{\delta}{\delta
f}\left[  \int\left\{  g^{\prime},E^{\prime}\right\}  f^{\prime}%
d\mathbf{\Gamma}^{\prime}\right]  \right\}  _{f=f_{o}}d\mathbf{\Gamma}%
\end{equation}
from the differential properties of the Poisson bracket we have
\begin{equation}
H^{\left(  2\right)  }=\frac{1}{2}\int f_{o}\left\{  g,\int\left[  \left\{
\frac{\delta g^{\prime}}{\delta f},E^{\prime}\right\}  f_{o}^{\prime}+\left\{
g^{\prime},\frac{\delta E^{\prime}}{\delta f}\right\}  f_{o}^{\prime}+\left\{
g^{\prime},E^{\prime}\right\}  \delta_{d}\left(  \mathbf{\Gamma}%
-\mathbf{\Gamma}^{\prime}\right)  \right]  d\mathbf{\Gamma}^{\prime}\right\}
d\mathbf{\Gamma}\;\;.
\end{equation}
An integration by parts of the two first terms in the middle sum and the
integration of the $\delta_{d}$ Dirac mass term gives
\begin{equation}
H^{\left(  2\right)  }=\frac{1}{2}\int f_{o}\left\{  g,\left[  \left\{
g,E\right\}  +\int\frac{\delta g^{\prime}}{\delta f}\left\{  E^{\prime}%
,f_{o}^{\prime}\right\}  d\mathbf{\Gamma}^{\prime}-\int\left\{  g^{\prime
},f_{o}^{\prime}\right\}  \frac{\delta E^{\prime}}{\delta f}d\mathbf{\Gamma
}^{\prime}\right]  \right\}  d\mathbf{\Gamma}\;\;.
\end{equation}
If $f_{o}$ is an equilibrium then $\left\{  E,f_{o}\right\}  $ vanishes, we
remark also that from $\left(  \ref{enepp}\right)  $
\begin{equation}
\frac{\delta E^{\prime}}{\delta f}=m\frac{\delta\psi^{\prime}}{\delta f}%
=\frac{-Gm^{2}}{\left|  \mathbf{q}-\mathbf{q}^{\prime}\right|  }\label{derpsi}%
\end{equation}
to arrive at
\begin{equation}
H^{\left(  2\right)  }=\frac{1}{2}\int f_{o}\left\{  g,\left\{  g,E\right\}
\right\}  d\mathbf{\Gamma}+\frac{Gm^{2}}{2}\int f_{o}\left\{  g,\int
\frac{\left\{  g^{\prime},f_{o}^{\prime}\right\}  }{\left|  \mathbf{q}%
-\mathbf{q}^{\prime}\right|  }d\mathbf{\Gamma}^{\prime}\right\}
d\mathbf{\Gamma}\;\;.
\end{equation}
A last integration by parts on each term allows us to see that
\begin{equation}
H^{\left(  2\right)  }=-\frac{1}{2}\int\left\{  g,f_{o}\right\}  \left\{
g,E\right\}  d\mathbf{\Gamma}-\frac{Gm^{2}}{2}\int\int\frac{\left\{
g,f_{o}\right\}  \left\{  g^{\prime},f_{o}^{\prime}\right\}  }{\left|
\mathbf{q}-\mathbf{q}^{\prime}\right|  }d\mathbf{\Gamma}^{\prime
}d\mathbf{\Gamma}%
\end{equation}

Studying of the sign of this quantity, one can study the stability of
equilibrium associated to $f_{o}$ against perturbations generated by $g$.
Classical results are presented in Perez and Aly\cite{Perezaly}, Perez et
al.\cite{perezetal} and Kandrup\cite{kandrup1}, \cite{Kandrup2}, \cite{Kandrup3}
. They can be summarized as follow :

\begin{itemize}
\item  If $f_{o}=f_{o}\left(  E\right)  $ and $f_{oE}:=\partial f_{o}/\partial
E<0$, then \ for all $g$ we have $H^{\left(  2\right)  }>0$. Hence, any
spherical and isotropic stellar system with an outwardly decreasing distribution function is linearly stable against all perturbations.

\item  If $f_{o}=f_{o}\left(  E,L^{2}\right)  $ , $f_{oE}<0$, and $g$ is such
that $\left\{  g,L^{2}\right\}  =0$ then $H^{\left(  2\right)  }>0$, such
perturbations are called preserving perturbations. All radial perturbations
are preserving\footnote{This proposition is not reciprocal.}. Hence, any
spherical and anisotropic stellar system with an outwardly decreasing distribution function is linearly stable at least against all radial perturbations.

\item  If $f_{o}=f_{o}\left(  E,L^{2}\right)  $ (resp. $f_{o}\left(
E,L_{z}\right)  )$,  $f_{oE}<0$, and $f_{oL^{2}}<0$ (resp. $f_{oL_{z}}<0$)
then there always exists non radial (resp. nonaxisymetric) perturbations,
corresponding to nonzero mass density perturbations, for which $H^{\left(
2\right)  }<0$. It was conjectured\cite{perezetal} that when $f_{o}\left(
E,L^{2}\right)  \rightarrow f_{o}\left(  E\right)  \delta_{d}\left(
L^{2}\right)  $ (resp. $f_{o}\left(  E,L_{z}\right)  \rightarrow f_{o}\left(
E\right)  \delta_{d}\left(  L_{z}\right)  $) then $H^{\left(  2\right)  }<0$
for all perturbation generator $g$. This may be associated to Radial Orbit Instability.
\end{itemize}

\subsection{Study of the transient regime of gravitational Vlasov-Poisson dynamics}

The transition to a steady state is a crucial problem in self-gravitating
systems dynamics. As a matter of fact, observations suggest that almost all
galaxies or globular clusters manifest overall similarities and that they do
not deviate very much from a statistical equilibrium. We may also note that
numerical simulation of collections of point masses tend to typically
evidence a rapid, systematic evolution, on dynamical time scale $T_{d}$,
towards a state, which is at least in a statistical sense, nearly
time-independent (see Roy and Perez\cite{Royetperez} ).

In the toy model of one dimensional gravity, where one considers \ the
interactions of infinite one dimensional-plane sheets, careful simulations can
be performed (e.g. Mineau et al.\cite{Mineauetal}). The net result of such
integrations is that typically such systems do not exhibit a pointwise
evolution toward a steady state configuration. It is significant that
solutions of $1D$ Vlasov-Poisson system for the same initial conditions also
exhibit an evolution toward a final state which is not strictly time-independent.

In the unphysical 2D gravity model, a complete thermodynamical approach can be
performed and show that there can exist an unique and well defined equilibrium
state associated to a maximum entropy principle (see Aly and
Perez\cite{alyperez} ). This result should be soon confronted with numerical
simulations of 2D Vlasov-Poisson system.

In the real 3D world, one can ask if there exists a correspondence
between time-dependent Vlasov-Poisson solutions, numerical simulations of large
collections of point masses and corresponding thermodynamical systems.\ 

In order to progress toward this end, let us present now, some uses of Hamiltonian
techniques presented above which perhaps could allow one to exhibit
time-dependent solutions of 3D\ Vlasov-Poisson system.

It is a fact that, at a given position $\mathbf{q}$ and for a given time $t$
the gravitational potential is a functional of the distribution function
\begin{equation}
\psi_{\mathbf{q},t}\left[  f\right]  =-Gm%
{\displaystyle\int}
\frac{f\left(  \mathbf{\Gamma}^{\prime},t\right)  }{\left|  \mathbf{q}%
-\mathbf{q}^{\prime}\right|  }d\mathbf{\Gamma}^{\prime}%
\end{equation}
one can then always take $F:=\psi$ in the functional Vlasov's equation to have
\begin{equation}
\frac{d\psi}{dt}=\left\langle \psi,H\right\rangle =-\left\langle
H,.\right\rangle \psi\label{psit}%
\end{equation}
hence, we can have a formal time development of the potential from which,
applying Laplacian operator one can deduce the temporal evolution of the mass
density. There is no problem in principle, rather only technical ones ...

As $-\left\langle H,.\right\rangle $ is a time-independent operator, we can
employ the resolvant of equation $\left(  \ref{psit}\right)  $ to write
\begin{align}
\psi_{\mathbf{q},t}\left[  f\right]    & =\psi_{\mathbf{q},t_{o}}^{(0)}%
+\tfrac{\left(  t-t_{o}\right)  }{1!}\psi_{\mathbf{q},t_{o}}^{(1)}%
+\tfrac{\left(  t-t_{o}\right)  ^{2}}{2!}\psi_{\mathbf{q},t_{o}}%
^{(2)}+\mathcal{O}\left(  \left(  t-t_{o}\right)  ^{2}\right)  \nonumber\\
& =\psi_{\mathbf{q},t_{o}}\left[  f_{o}\right]  -\tfrac{\left(  t-t_{o}%
\right)  }{1!}\left\langle H,\psi_{\mathbf{q},t_{o}}\left[  f_{o}\right]
\right\rangle _{f=f_{o}}\nonumber\\
& +\tfrac{\left(  t-t_{o}\right)  ^{2}}{2!}\left\langle H,\left\langle
H,\psi_{\mathbf{q},t_{o}}\left[  f_{o}\right]  \right\rangle \right\rangle
_{f=f_{o}}+\mathcal{O}\left(  \left(  t-t_{o}\right)  ^{2}\right)
\end{align}
The three first terms of this expansion are easily calculable.\\
The first term is obtained by definition
\begin{equation}
\psi_{\mathbf{q},t_{o}}\left[  f_{o}\right]  :=-Gm%
{\displaystyle\int}
\frac{f\left(  \mathbf{\Gamma}^{\prime},t_{o}\right)  }{\left|  \mathbf{q}%
-\mathbf{q}^{\prime}\right|  }d\mathbf{\Gamma}^{\prime}%
\end{equation}
From this relation one can see that
\begin{equation}
\frac{\delta\psi_{\mathbf{q},t_{o}}}{\delta f}=\frac{-Gm}{\left|
\mathbf{q}-\mathbf{q}^{\prime}\right|  }%
\end{equation}
and one can compute the second term
\begin{align}
\psi_{\mathbf{q},t_{o}}^{(1)}  & :=-\left\langle H,\psi_{\mathbf{q},t_{o}%
}\left[  f_{o}\right]  \right\rangle _{f=f_{o}}=-\int\left\{  \frac{\delta
H^{\prime}}{\delta f},\frac{\delta\psi_{\mathbf{q},t_{o}}}{\delta f}\right\}
f\left(  \mathbf{\Gamma}^{\prime},t_{o}\right)  d\mathbf{\Gamma}^{\prime
}\nonumber\\
& =\int\left\{  E^{\prime},Gm\left|  \mathbf{q}-\mathbf{q}^{\prime}\right|
^{-1}\right\}  f\left(  \mathbf{\Gamma}^{\prime},t_{o}\right)  d\mathbf{\Gamma
}^{\prime}\nonumber\\
& =-G\int\mathbf{p}^{\prime}.\frac{\partial\left|  \mathbf{q}-\mathbf{q}%
^{\prime}\right|  ^{-1}}{\partial\mathbf{q}^{\prime}}f\left(  \mathbf{\Gamma
}^{\prime},t_{o}\right)  d\mathbf{\Gamma}^{\prime}%
\end{align}
A first hypothesis on the initial distribution function can be now posed to
simplify the calculus.

Hypothesis 1 : The initial state is symmetric in the velocity space. The
consequence of this assumption is that for any vectorial space field $B\left(
\mathbf{q}\right)  $ and any vectorial \emph{odd} velocity field $A\left(
\mathbf{p}\right)  ,$ $\ $\ we have
\begin{equation}
\int A\left(  \mathbf{p}\right)  .B\left(  \mathbf{q}\right)  \;f\left(
\mathbf{\Gamma},t_{o}\right)  d\mathbf{\Gamma}=0
\end{equation}
Under this non-restrictive assumption, one can easily show that
\begin{equation}
\psi_{\mathbf{q},t_{o}}^{(1)}=0
\end{equation}
Moreover, in a general way one see also that
\begin{equation}
\frac{\delta\psi_{\mathbf{q},t_{o}}^{(1)}}{\delta f}=-G\mathbf{p}^{\prime
}.\frac{\partial\left|  \mathbf{q}-\mathbf{q}^{\prime}\right|  ^{-1}}%
{\partial\mathbf{q}^{\prime}}%
\end{equation}
which allows the computation of the third term in the potential expansion.
\begin{align}
\psi_{\mathbf{q},t_{o}}^{(2)}  & =\left\langle H,\psi_{\mathbf{q},t_{o}}%
^{(1)}\right\rangle _{f=f_{o}}=-G\int\left\{  E^{\prime},\mathbf{p}^{\prime
}.\frac{\partial\left|  \mathbf{q}-\mathbf{q}^{\prime}\right|  ^{-1}}%
{\partial\mathbf{q}^{\prime}}\right\}  f\left(  \mathbf{\Gamma}^{\prime}%
,t_{o}\right)  d\mathbf{\Gamma}^{\prime}\\
& =\mathcal{A}+\mathcal{B}\nonumber
\end{align}
where
\begin{align}
\mathcal{A}  & =-Gm\int d\mathbf{\Gamma}^{\prime}f\left(  \mathbf{\Gamma
}^{\prime},t_{o}\right)  \dfrac{\partial\psi_{\mathbf{q}^{\prime},t_{o}}%
}{\partial\mathbf{q}^{\prime}}.\dfrac{\partial\left|  \mathbf{q}%
-\mathbf{q}^{\prime}\right|  ^{-1}}{\partial\mathbf{q}^{\prime}}%
\;\;\label{B}\\
& \text{and}\;\;\;\nonumber\\
\mathcal{B}  & =+\frac{G}{m}\int d\mathbf{\Gamma}^{\prime}f\left(
\mathbf{\Gamma}^{\prime},t_{o}\right)  \mathbf{p}^{\prime}.\dfrac{\partial
}{\partial\mathbf{q}^{\prime}}\left(  \mathbf{p}^{\prime}.\dfrac
{\partial\left|  \mathbf{q}-\mathbf{q}^{\prime}\right|  ^{-1}}{\partial
\mathbf{q}^{\prime}}\right)
\end{align}
$\psi_{\mathbf{q},t_{o}}^{(2)}$ could be greatly simplified assuming some
physical properties of the initial state $f\left(  \mathbf{\Gamma}%
,t_{o}\right)  $. First, we suppose that the initial density is constant on
a bounded domain $\Omega\subset\mathbb{R}^{3}$. Then we may pose

Hypothesis 2
\begin{equation}
\int f\left(  \mathbf{\Gamma},t_{o}\right)  d\mathbf{p=}\nu_{o}.\chi_{\Omega
}=\frac{\rho_{o}}{m}.\chi_{\Omega}%
\end{equation}
where $\rho_{o}=m\nu_{o}$ is a strictly positive constant ($\nu_{o}$ is the
value of the uniform initial number density whereas $\rho_{o}$ represent the
mass density) and $\chi_{\Omega}$ denotes the characteristic
function\footnote{Classically, the characteristic function is defined to be
unity on $\Omega$, and is zero elsewhere.} on $\Omega$.

This hypothesis which can appear very restrictive, is fairly adapted for the
description of the initial state of the physical material before it collapse
to form a self-gravitating structure. Under this hypothesis, we have
\begin{equation}
\mathcal{A}=-G\rho_{o}\int_{\mathbf{q}\prime\in\Omega}\dfrac{\partial
\psi_{\mathbf{q}^{\prime},t_{o}}}{\partial\mathbf{q}^{\prime}}.\dfrac
{\partial\left|  \mathbf{q}-\mathbf{q}^{\prime}\right|  ^{-1}}{\partial
\mathbf{q}^{\prime}}\;\;,
\end{equation}
an integration by parts in which all surface terms vanish then gives
\begin{equation}
\mathcal{A}=G\rho_{o}\int_{\mathbf{q}\prime\in\Omega}\psi_{\mathbf{q}%
^{\prime},t_{o}}.\Delta^{\prime}\left(  \left|  \mathbf{q}-\mathbf{q}^{\prime
}\right|  ^{-1}\right)\;\;.
\end{equation}
It is well known that $\left|  \mathbf{q}-\mathbf{q}^{\prime}\right|  ^{-1}$
is proportional to the Fundamental solution of the 3D radial Laplacian, more
precisely we have
\begin{equation}
\Delta^{\prime}\left(  \left|  \mathbf{q}-\mathbf{q}^{\prime}\right|
^{-1}\right)  =-4\pi\delta_{d}\left(  \mathbf{q}-\mathbf{q}^{\prime}\right)
\end{equation}
hence
\begin{equation}
\mathcal{A}=-\frac{\psi_{\mathbf{q},t_{o}}}{\tau_{o}^{2}}%
\end{equation}
where we have introduced the Jeans time $\tau_{o}:=\left(  4\pi G\rho
_{o}\right)  ^{-1/2}$.

The second term is a little bit more complicated, in Cartesian coordinates we
have
\begin{equation}
\mathcal{B}=\frac{G}{m}\sum_{i,j=1}^{3}\int d\mathbf{\Gamma}^{\prime}f\left(
\mathbf{\Gamma}^{\prime},t_{o}\right)  p_{i}^{\prime}p_{j}^{\prime}%
\dfrac{\partial^{2}\left|  \mathbf{q}-\mathbf{q}^{\prime}\right|  ^{-1}%
}{\partial q_{i}^{\prime}\partial q_{j}^{\prime}}%
\end{equation}
We now make our last assumption.

Hypothesis 3 : The initial state has an isotropic and constant velocity
dispersion $\sigma^{2}$%
\begin{equation}
\forall i=1,2,3\;\;\int d\mathbf{p}\text{ }f\left(  \mathbf{\Gamma}%
,t_{o}\right)  p_{i}^{2}=\;\nu_{o}\frac{m^{2}\sigma^{2}}{3}\chi_{\Omega}%
=\rho_{o}\frac{m\sigma^{2}}{3}\chi_{\Omega}%
\end{equation}

Physically this assumption means that the initial state is isothermal.

Coupling all the hypotheses, we have
\begin{equation}
\mathcal{B}=\sigma^{2}G\rho_{o}\int_{\mathbf{q}\prime\in\Omega}\Delta^{\prime
}\left(  \left|  \mathbf{q}-\mathbf{q}^{\prime}\right|  ^{-1}\right)
=-\frac{\sigma^{2}}{\tau_{o}^{2}}%
\end{equation}
The beginning of the potential expansion may be written as
\begin{equation}
\psi_{\mathbf{q},t}\left[  f\right]  =\psi_{\mathbf{q},t_{o}}\left[
f_{o}\right]  -\tfrac{\left(  t-t_{o}\right)  ^{2}}{2\tau_{o}^{2}}\left[
\psi_{\mathbf{q},t_{o}}\left[  f_{o}\right]  +\sigma^{2}\right]
+\mathcal{O}\left(  \left(  t-t_{o}\right)  ^{2}\right)
\end{equation}
Attempting to go further in this expansion is extremely perilous. As a matter of fact, we
see from the term $\mathcal{A}$ that, the functional derivatives of $\psi
_{\mathbf{q},t_{o}}^{(2)}$ do not come directly due to the $f-$non linearity
incoming in the term $\psi.\,f$. The calculus then becomes extremely tedious ... 

At this stage, only some general remarks can be made using such a technique in
a general way :

\begin{itemize}
\item  Under hypothesis 2, all odd terms vanish in the expansion
\begin{equation}
\psi_{\mathbf{q},t_{o}}^{(2k+1)}=0\;,\;\;\forall k\in\mathbb{N}
\end{equation}

\item  Keeping only linear terms, it seems\footnote{It is more conjectured
using preliminar results from high order terms, than strictly proved ...} that
only terms of the same nature of \ $\psi_{\mathbf{q},t_{o}}^{(2)}$ remains in
the expansion, i.e.
\begin{equation}
\psi_{\mathbf{q},t_{o}}^{(2k)}\propto  (-1)^k \frac
{\psi_{\mathbf{q},t_{o}}\left[  f_{o}\right]  }{\tau_{o}^{2k}}+cst\;,\;\;\forall k\in\mathbb{N}
\end{equation}
neglecting the constant, this gives
\begin{equation}
\psi_{\mathbf{q},t}=\psi_{\mathbf{q},t_{o}}\left[  f_{o}\right]  \sum
_{k=0}^{\infty}(-1)^k\frac{\left(  t-t_{o}\right)  ^{2k}}{\left(  2k\right)
!\tau_{o}^{2k}}=\psi_{\mathbf{q},t_{o}}\left[  f_{o}\right]  
\cos\left( \frac{t-t_{o}}{\tau_{o}}\right)
\end{equation}
this maybe corroborates  the $\tau_{o}-$periodic damped regime 
which appear in generic gravitational collapse numerical experiments.
\end{itemize}
 We stress that there are preliminary results. We hope to give a more complete description of the transient regime of Vlasov-Poisson dynamics in an upcoming article.

\end{document}